\documentclass[superscriptaddress,twocolumn,nofootinbib]{revtex4}
\usepackage{bm,graphicx}
\usepackage{hyperref}

\begin{document}

\title{On derivation of Dresselhaus spin-splitting
Hamiltonians in one-dimensional electron systems}

\author{I.A. Kokurin}
\email[E-mail:]{kokurinia@math.mrsu.ru} \affiliation{Institute of
Physics and Chemistry, Mordovia State University, 430005 Saransk,
Russia} \affiliation{Ioffe Institute, 194021 St. Petersburg, Russia}
\affiliation{St. Petersburg Electrotechnical University ``LETI'',
197022 St. Petersburg, Russia}

\begin{abstract}
Two-dimensional (2D) semiconductor structures of materials without
inversion center (e.g. zinc-blende ${\rm A^{III}B^V}$) possess the
zero-field conduction band spin-splitting (Dresselhaus term), which
is linear and cubic in wavevector $k$, that arises from cubic in $k$
splitting in a bulk material. At low carrier concentration the cubic
term is usually negligible. However, if we will be interested in the
following dimensional quantization (in 2D plane) and the character
width in this direction is comparable with the width of
2D-structure, then we have to take into account $k^3$-terms as well
(even at low concentrations), that after quantization leads to
comparable contribution that arises from $k$-linear term. We propose
the general procedure for derivation of Dresselhaus spin-splitting
Hamiltonian applicable for any curvilinear 1D-structures. The simple
examples for the cases of a quantum wire (QWr) and a quantum ring
(QR) defined in usual [001]-grown 2D-structure are presented.
\end{abstract}

\date{\today}

\maketitle

\section{Introduction}

The spin-orbit splitting of conduction band plays a key role in
various effects, such as: (i) Dyakonov-Perel spin
relaxation~\cite{Dyakonov1972,Dyakonov1976}; (ii) photo-galvanic
effects~\cite{Ivchenko2008}, (iii) current-induced spin polarization
(see for instance recent overview~\cite{Averkiev2017} and references
therein); (iv) intrinsic spin Hall effect~\cite{Sinova2004} and some
others. In low-dimensional structures there are three principal
mechanisms of spin-orbit coupling (SOC): (i) structure inversion
asymmetry (SIA) or Rashba term~\cite{Bychkov1984} that is due to
asymmetry of a heterostructure potential profile; (ii) bulk
inversion asymmetry (BIA) or Dresselhaus term~\cite{Dyakonov1976}
arising due to the lack of inversion center in a bulk
material~\cite{Dresselhaus1955}; and (iii) interface inversion
asymmetry (IIA) leading to a renormalization of the BIA
contribution~\cite{Rossler2002}. Here we will discuss BIA term in 1D
system defined in 2D-structure by electrostatic gating or etching,
e.g. quantum wire (QWr), quantum ring (QR) or more complex
curvilinear 1D structures.

The conduction band spin-splitting in bulk ${\rm A^{III}B^V}$
material is cubic in momentum, and it is given
by~\cite{Dresselhaus1955}
\begin{equation}
\label{H_D_bulk} H_{\rm BIA}=\gamma\bm{\sigma\kappa},
\end{equation}
where $\bm{\sigma}=(\sigma_x,\sigma_y,\sigma_z)$ is the vector of
the Pauli matrices, $\kappa_x=k_x(k_y^2-k_z^2)$, and $\kappa_y$,
$\kappa_z$ can be found by cyclic permutations. It should be noted,
that such a form of Dresselhaus Hamiltonian is true for the
reference frame with $x||[100]$, $y||[010]$, $z||[001]$.

The Dresselhaus SOC for quantum well (QW) can be found from
Eq.~(\ref{H_D_bulk}) by averaging on a wave-function $\psi_0(z)$ of
a ground state of the quantized motion
\begin{equation}
\label{H_D_2D} H_{\rm BIA}^{\rm
2D}=\beta_z(\sigma_yk_y-\sigma_xk_x)+\gamma(\sigma_xk_xk_y^2-\sigma_yk_yk_x^2),
\end{equation}
where $\beta_z=\gamma\langle k_z^2\rangle$, with $\langle
k_z^2\rangle=\langle\psi_0|k_z^2|\psi_0\rangle$. Frequently, at low
concentration due to inequality $k_{x,y}\ll \langle
k_z^2\rangle\simeq (\pi/w_z)^2$ only the first $k$-liner term in
Eq.~(\ref{H_D_2D}) is taken into account, with $w_z$ being the width
of 2D-layer. These results are derived for [001]-grown QW.

If a cubic in $k$ term in Eq.~(\ref{H_D_2D}) is neglected, then it
leads to some misunderstanding when one tries to derive Dresselhaus
Hamiltonian for the structure of the lower dimension based on 2D
one, e.g. a quantum wire. The strict procedure means the use of the
bulk Dresselhaus term (\ref{H_D_bulk}) as a starting point with the
quantization (and corresponding averaging) in two directions. For
QWr defined in [001]-grown 2D-system and with transport direction
along [100] this leads to

\begin{equation}
H_{\rm BIA}^{\rm 1D}=(\beta_y-\beta_z)\sigma_xk_x,
\end{equation}
where $\beta_y=\gamma\langle k_y^2\rangle\simeq\gamma(\pi/w_y)^2$
with $w_y$ being the QWr width in the plane of initial 2D-structure.
The use of $k$-linear term only as staring point leads in this case
to the same result, but with $\beta_y=0$. This is true for QWrs
which width $w_y$ is much wider than $w_z$, but the condition of the
single subband filling (one-dimensionality) holds. Thus, the result
that is obtained from the $k$-linear operator differs only by
strength parameter. But it is worth noting, that in QWr of the
square cross-section there is no spin-splitting due to
$\beta_z=\beta_y$. It means that the $k$-linear Dresselhaus term is
not enough, especially good it can be seen for structures with other
crystallographic orientation, both the orientation of initial
2D-structure and the QWr orientation in 2D plane.

\begin{figure*}
\includegraphics[width=0.7\linewidth]{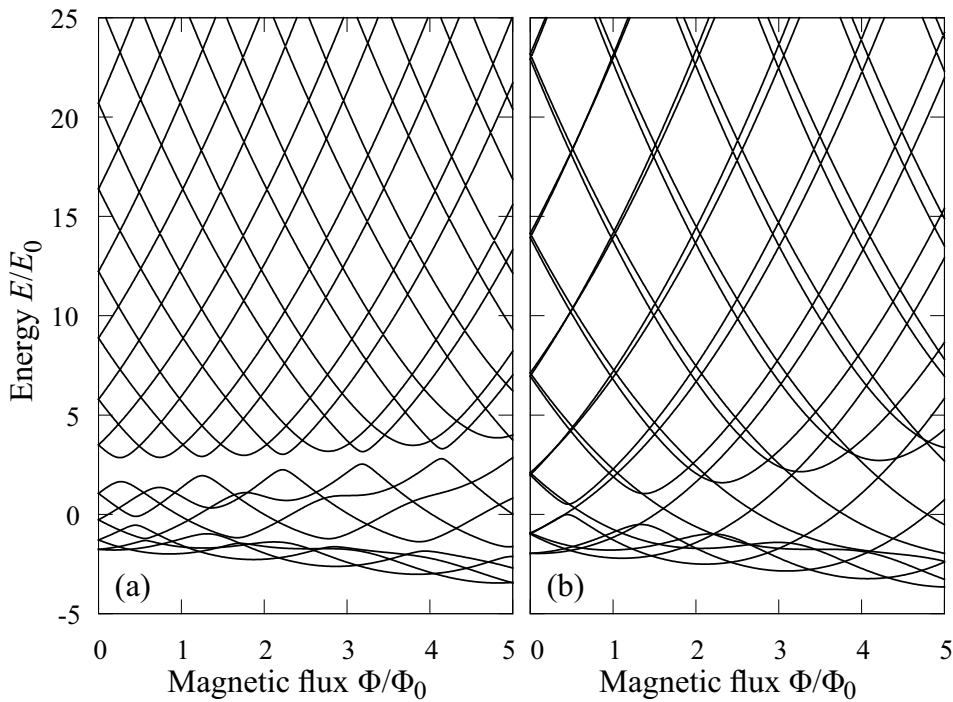}
\caption{The energy levels of QR with Dresselhaus SOC as a function
of magnetic flux. Parameters are following: $g=12.0$,
$m^*/m_0=0.05$, $2\beta_zm^*r_0/\hbar^2=2.8$,
$E_0=\hbar^2/2m^*r_0^2$. (a) Total BIA contribution are taken into
account, $\beta_r=\beta_z/2$, (b) analytical result for the
simplified model with $\beta_r=0$.}
\end{figure*}

\section {General approach for derivation of BIA Hamiltonian in one dimension}

Using the simple example of QWr we shown that in general the
$k$-linear SOC Hamiltonian is poor starting point at dimension
lowering. Method of invariants~\cite{Ivchenko1997} permits to find
the form of BIA Hamiltonians from symmetry consideration for certain
structure with defined crystallographic orientation. However, the
strength of entering parameters and the relationship between them
can be found only by means of strict procedure starting from bulk
BIA-splitting. But for curvilinear structures, such as QR, the
strict procedure based on a ground-state averaging, is much more
complicated.

We propose the method of the effective spin-orbit Hamiltonian
derivation based on our previous work~\cite{Kokurin2015}, where we
used a simple approach. Our approach includes the following steps:
(i) the derivation of bulk BIA Hamiltonian for necessary
crystallographic orientation using the coordinate transformation;
(ii) the dimension lowering (to 2D) of a bulk Hamiltonian (an
averaging on a ground-state wave function of the transverse motion,
$z$-direction) with conservation of high-power terms in $k$; (iii)
an additional rotation of coordinates (around $z$-axis on $\varphi$
angle) that with simultaneous quantization (averaging) in new
$x$-direction leads to BIA Hamiltonian for 1D QWr that orientation
coincides with the tangential direction to point with curvilinear
coordinates $(\xi,\zeta)$; (iv) the replacement
$\sigma_x\rightarrow\sigma_\xi$, $\sigma_y\rightarrow\sigma_\zeta$,
$k_y\rightarrow k_\zeta$ with the following symmetrization of
non-commuting terms (i.e. $AB\rightarrow \{A,B\}=(AB+BA)/2$), since
Pauli matrices $\sigma_\xi$ and $\sigma_\zeta$ in a `new' reference
frame in general do not commute with remaining momentum operator
$k_\zeta$.

\section{BIA Hamiltonian for thin quantum ring}

Let us apply above scheme to the derivation of BIA Hamiltonian for
thin (1D) circular QR of radius $r_0$ defined in a usual
2D-structure with [001]-orientation. Two first points in our scheme
can be omitted, and we start from Hamiltonian (\ref{H_D_2D}). After
rotating of coordinates and $x$-quantization we have to replace
$\sigma_x$ and $\sigma_y$ by
$\sigma_r(\varphi)=\cos\varphi\sigma_x+\sin\varphi\sigma_y$,
$\sigma_\varphi(\varphi)=-\sin\varphi\sigma_x+\cos\varphi\sigma_y$,
respectively. Remaining momentum $k_y$ has to be replaced by
$K_\varphi=r_0^{-1}(-i\partial/\partial\varphi+\Phi/\Phi_0)$, where
we include in consideration the magnetic field ${\bf B}$ with the
flux $\Phi=\pi r_0^2B$ and flux quantum $\Phi_0=2\pi\hbar c/|e|$.
After symmetrization we find
\begin{eqnarray}
\label{H_BIA} \nonumber H_{\rm BIA}^{\rm ring}=(\beta_z-\beta_r)\cos
2\varphi\sigma_\varphi
K_\varphi+\left(\beta_z+\frac{\beta_r}{2}\right)\sin
2\varphi\sigma_r
K_\varphi\\
+\left(\beta_z-\frac52\beta_r\right)\frac{i}{2r_0}\sin
2\varphi\sigma_\varphi-(\beta_z+2\beta_r)\frac{i}{2r_0}\cos
2\varphi\sigma_r,
\end{eqnarray}
where $\beta_r\simeq\gamma(\pi/w_r)^2$, with $w_r$ being the radial
width of the ring.

If the QR radial width is much wider than its axial width ($w_r\gg
w_z$) then the inequality $\beta_z\gg\beta_r$ is fulfilled and
neglecting terms containing $\beta_r$ in Eq.~(\ref{H_BIA}) we find
after some algebra the following operator
\begin{equation}
H=\beta_z\left[\sigma_\varphi(-\varphi)K_\varphi-\frac{i}{2r_0}\sigma_r(-\varphi)\right],
\end{equation}
that was derived in Ref.~\cite{Sheng2006} taking into account only
$k$-linear terms in BIA SOC of initial 2D Hamiltonian.

Now consider the total Hamiltonian containing a kinetic term
$\hbar^2K_\varphi^2/2m^*$ ($m^*$ is the electron effective mass),
the Zeeman splitting $H_Z=(1/2)g\mu_BB\sigma_z$ ($g$ is the
effective g-factor and $\mu_B$ is the Bohr magneton) and BIA SOC
from Eq.~(\ref{H_BIA}). Such a Hamiltonian can be diagonalized only
numerically. One can utilize any appropriate basis set, e.g.
$\Psi_{ms}(\varphi)=(2\pi)^{-1/2}e^{im\varphi}\chi_s$, with $m=0,
\pm 1, \pm 2,...$ and
$\chi_{+1}=\left(\begin{array}{c}1\\0\end{array}\right)$,
$\chi_{-1}=\left(\begin{array}{c}0\\1\end{array}\right)$. The
results of a numerical diagonalization are depicted in Fig.~1a. The
reduced Hamiltonian matrix 102$\times$102 ($m=0,\pm 1, ..., \pm 25$
and $s=\pm 1$) gives perfect precision for plotted levels. One can
see a significant difference in Fig.~1b, where the simplified BIA
term was used ($\beta_r=0$)~\cite{Sheng2006}. In this case the
spectral problem can be solved analytically due to the fact that
Hamiltonian commutes with the following operator,
$f_z=-i\hbar\partial/\partial\varphi-(\hbar/2)\sigma_z$.

\section{Conclusions}

In conclusion, we have shown that the use of $k$-linear 2D BIA-term
as a starting point is not enough for derivation of correct 1D BIA
Hamiltonian. We proposed the approach of the SOC Hamiltonian
derivation for any 1D structure that defined in a 2D-system from a
material without the inversion center. The BIA Hamiltonian for thin
QR oriented in (001)-plane is constructed. This Hamiltonian can be
used to calculate the persistent charge and spin currents using the
equilibrium density matrix formalism~\cite{Kokurin2018}.

\end{document}